# Asymmetric disease dynamics in multihost interconnected networks

Shai Pilosof[*1,2], Gili Greenbaum[*2,3], Boris R. Krasnov[2], and Yuval R. Zelnik[3]

[1]*Department of Ecology and Evolution, University of Chicago, 1103 E 57 st, Chicago, 60637, USA*

[2]*Mitrani Department of Desert Ecology, Blaustein Institutes for Desert Research, Ben-Gurion University of the Negev, Sede Boqer Campus, 84990, Israel*

[3]*Department of Solar Energy and Environmental Physics, Blaustein Institutes for Desert Research, Ben-Gurion University of the Negev, Sede Boqer Campus, 84990, Israel*

Corresponding author: Pilosof, S. (pilosofs@uchicago.edu)

---

[*]Contributed equally.

# Abstract

Epidemic spread in single-host systems strongly depends on the population's contact network. However, little is known regarding the spread of epidemics across networks representing populations of multiple hosts. We explored cross-species transmission in a multilayer network where layers represent populations of two distinct hosts, and disease can spread across intralayer (within-host) and interlayer (between-host) edges. We developed an analytic framework for the SIR epidemic model to examine the effect of (i) source of infection and (ii) between-host asymmetry in infection probabilities, on disease risk. We measured risk as outbreak probability and outbreak size in a focal host, represented by one network layer. Numeric simulations were used to validate the analytic formulations. We found that outbreak probability is determined by a complex interaction between source of infection and between-host infection probabilities, whereas outbreak size is mainly affected by the non-focal host to focal host infection probability alone. Hence, inter-specific asymmetry in infection probabilities shapes disease dynamics in multihost networks. These results expand current theory of monolayer networks, where outbreak size and probability are considered equal, highlighting the importance of considering multiple measures of disease risk. Our study advances understanding of multihost systems and non-biological systems with asymmetric flow rates.

# 1 Introduction

Network models are commonly used to study disease transmission in human and animal populations because they consider heterogeneities in the contact structure of populations [1–3]. However, most network models assume that the network is not connected to other networks, while in the real world, isolated networks are the exception rather than the rule. For example, in humans, diseases can spread through interconnected transportation networks, or through interconnected social groups.

One way to model disease spread across several networks is with multilayer networks, in which different layers can represent different populations. In particular, interconnected networks (as defined by [4]) are a useful representation because they contain two types of edges: *intralayer edges* connect individuals from the same population, while *interlayer edges* connect individuals from different populations (Fig. 1). Disease transmission in interconnected networks has been explored in the field of physics, and driven by human-related examples (reviewed in [4–6]). The focus of virtually all previous studies has been on patterns of interlayer connectivity and the distribution of intralayer vs. interlayer edges in the network, which is the main characteristic differentiating interconnected systems from single-network systems [4–6]. For example, Saumell-Mendiola *et al.* [7] have shown that when the correlation between the intralayer and interlayer degree distributions is strong, an outbreak may occur in the system even if it would not have occurred in any of the single layers alone.

Despite the focus on human-related systems, it is important to note the tremendous effects of pathogen transmission in interconnected non-human systems. In nature, populations of the same or different species interconnect in ways that allow for pathogen exchange, including predation [8], sharing space [9, 10] and sharing food sources [11]. In particular, cross-species transmission has attracted much attention due to its possible impact on agriculture (e.g., transmission between wildlife and domestic animals), persistence of wild populations, and species conservation efforts [12–15]. For example, white-nose syndrome, induced by the fungus *Pseudogymnoascus destructans*, which invaded North America from Europe, is causing major declines in bat populations of several species [15, 16]. From a public health perspective, cross-species transmission of zoonotic diseases between animals and humans has been a major focus in disease ecology [17, 18], with notable examples such as avian influenza and Ebola [19, 20]. However, to date, cross-species pathogen transmission has not been studied within a network analysis framework [13, 21–23] (but see [24]).

Here, we bring together the theoretical aspects of both the physics of infection processes on interconnected networks and of the disease ecology of multihost systems. We adopt an ecological point of view according to which an isolated network represents a population of a particular host species while an interconnected network represents a multihost system composed of populations of distinct host species (Fig. 1). This view is both realistic and necessary because (i) in multihost systems one host can alter the dynamics of pathogens in other hosts [13, 21–23] and (ii) these dynamics can be affected by the underlying network structure of each host.

In disease ecology, one type of multihost pathogen transmission models focuses on the case in which a target species is infected by a source species [13, 25]. In this type of models, little or no transmission from the target back to the source is assumed [22, 25]. A second kind of models deals with diseases that can be transmitted and maintained by more than one species (e.g., bovine tuberculosis or canine distempr virus). In this case, the dynamics of disease in the host of interest, which we term the *focal host*, may be affected by two critical factors inherent to multihost systems: (i) the source of infection—if the disease originates in the focal host itself or in a non-focal host; and (ii) the asymmetry in the rate of transmission between the two hosts. Both of these factors are of crucial importance for cross-species pathogen transmission [9, 13, 22]. For example, recurrent infections from a non-focal host can cause endemic infection in a focal host even if the pathogen cannot establish in it [13]. Additionally, in zoonotic diseases the source of infection is the animal, rather than the human, causing strong asymmetry in infection dynamics; that is, probability of infection is higher from an animal species to humans than the other way around [18, 26].

Our aim is to understand how the interaction between these two factors affects disease dynamics in an interconnected network system. We quantify dynamics using two measures: (i) The probability of an outbreak (i.e., a significant portion of the population is infected). (ii) The expected size of an outbreak (i.e., the proportion of the population infected), when an outbreak occurs. We address this aim by developing a novel analytic framework to quantify outbreak size and probability in interconnected networks, and by simulating disease spread.

## 2 Modelling pathogen spread in interconnected networks

Following previous studies on interconnected networks [7, 27, 28], we use interconnected networks as depicted in Fig. 1. We refer to each of the single networks in an interconnected network as *layers* [4]. *Intralayer* edges connect nodes within a layer while *interlayer* edges connect nodes from different layers. For simplicity, we considered the case of two interconnected populations (belonging to different hosts). We explore disease dynamics in layer A ($L_A$) of the interconnected network and thus consider $L_A$ as our *focal host* species and layer B ($L_B$) as the *non-focal species* (Fig. 1). We define the mean number of interlayer edges connected to a node (or 'mean interlayer degree') in $L_A$ as $e_A = \frac{E}{n_A}$, where $E$ is the number of interlayer edges and $n_A$ is the number of nodes in $L_A$. The mean interlayer degree of a node in $L_B$ is analogously defined as $e_B = \frac{E}{n_B}$.

We study the spread of a pathogen in interconnected networks with an SIR model, in which each individual belongs to one of three compartments: susceptible (S), infected and thus infectious (I) or resistant and not infectious (R). It is advantageous to work with the SIR model because it is relevant for a vast range of diseases and because it is well established in the network epidemiology literature [3], providing us with a sound theoretical basis to build upon.

Following [29], we denote the probability that a contact between an infectious and a susceptible individual leads to successful transmission of infection in a given time step as $\beta$ (this is the equivalent of $\nu$ from Begon *et al.* [29]). We hereafter refer to this parameter as *infection probability*. Because nodes in different layers belong to different species, the probability that a susceptible individual will be infected by an infectious neighbour is determined by the species identity of both. Hence, $\beta$ depends on the layers to which the two nodes belong [22]. We thus defined the infection matrix

$$\beta = \left[ \begin{array}{cc} \beta_{AA} & \beta_{AB} \\ \beta_{BA} & \beta_{BB} \end{array} \right] \quad (1)$$

For example, $\beta_{AB}$ is the probability that a node in $L_A$ will be infected by a node from $L_B$. Individuals move from an infected state to a recovered state after a given amount of time steps (the infectious period, $\tau$), and the recovery rate is therefore $\gamma = \frac{1}{\tau}$ (sensu [30]).

We use the SIR model to examine two infection scenarios. In Scenario 1 the epidemic originates with an individual in $L_A$ (the focal host) whereas in Scenario 2 it originates with an individual in $L_B$ (the non-focal host). In both scenarios we track only the population of the focal host.

# 3 Analytic formulation of outbreak size and probability

The SIR spreading process on a network can be analytically studied by equating it with a bond percolation process [1, 31]. The bond percolation problem concerns diffusion through a discrete substrate to form clusters. In monolayer (non interconnected) networks, the probability of a large outbreak and the expected size of such an outbreak in arbitrary random networks (random networks with any degree distribution) has been described by Newman [32]. A percolation process on a network of size $n$ may be subcritical, in which case the diffusion occurs only in small clusters or components, which are of size in the order of $log(n)$. Alternatively, it can be supercritical, in which case a *giant component* emerges, with size in the order of $n$, and possibly other small non-giant components. An outbreak can occur only when the system is in the supercritical phase; in epidemiological terms, this means that the disease crosses the $R_0 > 1$ threshold. Because we are interested in those settings where an epidemic may potentially have serious consequences for populations, we focus on scenarios in the supercritical phase.

The *transmissibility* of the pathogen via a given edge depends on both the infection probability and the recovery rate and is $T = 1 - (1-\beta)^{\frac{1}{\gamma}}$. This property is a measure of the likelihood that the disease will be transmitted via a given edge if one of the nodes adjacent to it is infected. Transmissibility allows evaluating the size of an outbreak, if one occurs, by estimating the expected size of the giant component (i.e., the expected fraction of the network occupied by the giant component) in the percolation process on the network [31, 32]. For a large random network with an arbitrary degree distribution, both the size of the giant component, $S$, and the mean size of the non-giant components, $s$, can be calculated using percolation theory (see Supplementary Information). Thus, an outbreak occurs when the epi-demic originates in an individual in the giant component and spreads to the entire giant component. The probability that an epidemic starting in a random individual will result in an outbreak, $P$, is the same as the probability of belonging to the giant component, and thus in monolayer networks $P = S$ [31, 32].

Applying percolation theory to interconnected networks is not straightforward, particularly when the infection probability is asymmetric ($\beta_{AB} \neq \beta_{BA}$). While the percolation process can describe the SIR dynamics in each of the layers separately (as for monolayer networks), the pathogen spreads across the interlayer edges with different probabilities in each direction. Therefore, there is no single transmissibility value for the interlayer edges, and the problem of finding the outbreak probability and size cannot be formulated as a simple percolation process. To provide approximation for outbreak size and probability in interconnected networks, we start by considering the percolation processes in each layer independently, as if they were disconnected. On top of that, we add the effect of the other layer for increasing the probability and size of outbreaks.

We assume (Assumption 1) that an epidemic in one layer will inevitably lead to an epidemic in the second layer, since the layers are sufficiently connected. This means that there are enough interlayer edges connecting the layers, and that the interlayer transmission probabilities are not too close to zero. Next, we consider the nature of the multihost setting. First, due to behavioral and life history differences between species, within-host contact rates are usually greater than between-host contact rates. Second, once two individuals come in contact, within-host infection probabilities are larger than within-host infection probabilities due to physiological competence of the host and the pathogen [13, 21, 33]. With this in mind, we assume (assumption 2) that one interlayer edge at most is expected to transmit the disease to or from each non-giant component.

## 3.1 Probability of outbreak in interconnected networks

We wish to find the analytic solution to $P_A$ — the probability of an outbreak in layer $L_A$. The probability of an outbreak in each of the layers if the interlayer edges are epidemiologically disconnected (i.e., $\beta_{AB} = 0$ and $\beta_{BA} = 0$) is the same as the size of the giant component in the percolation process in these layers; we denote these sizes as $S_A$ and $S_B$. We further denote $s_A$ and $s_B$ as the mean number of nodes in the non-giant components in layers $A$ and $B$, respectively, when the layers are disconnected. Note that $S_A$ and $S_B$ are measured as a fraction of the network size, while $s_A$ and $s_B$ are measured in absolute number of nodes, rather than in fractions. When the layers are connected, the transmissibility of the pathogen from $L_A$ to $L_B$ via the interlayer edges is $T_{AB} = 1 - (1 - \beta_{AB})^{\frac{1}{\gamma_A}}$, where $\gamma_A$ is the recovery rate of the host in $L_A$. This gives the probability that the pathogen spreads from an infected individual in $L_A$ to a connected individual in $L_B$ through a given interlayer edge. $T_{BA}$ is analogously defined as $T_{BA} = 1 - (1 - \beta_{BA})^{\frac{1}{\gamma_B}}$.

In Scenario 1, an outbreak may occur as a result of intralayer transmissions within $L_A$ if the node of origin belongs to the giant component. Alternatively, if the node of origin is not in the giant component, an epidemic may occur as a result of transmission to $L_B$ causing an outbreak there which is then transmitted back to $L_A$. This will occur if (i) the non-giant component in which the disease originated is connected to $L_B$ via an interlayer edge (with expected value $e_A s_A$). (ii) The interlayer edge transmits the disease to $L_B$ (with probability $T_{AB}$). (iii) The node to which the pathogen is transmitted is in the giant component in $L_B$ (with probability $S_B$). Therefore, in Scenario 1

$$P_A = S_A + (1 - S_A) S_B e_A s_A T_{AB}. \qquad (2)$$

The first term in the equation refers to the case in which disease orig-inates in the giant component whereas the second term refers to the concurrence of the three conditions mentioned, in the case where the disease originates in the non-giant component. We see that the probability of an outbreak depends only on the interlayer edge transmissibility in the direction $L_A \to L_B$. This is so because the infection in the other direction will occur if there is an outbreak in $L_B$, regardless of the transmissibility of the epidemic in the direction $L_B \to L_A$ (due to assumption 1).

In scenario 2, the disease originates in $L_B$ and the logic is analogous. An outbreak will occur in $L_A$ if an outbreak occurs in $L_B$. Alternatively, if an outbreak does not occur in $L_B$, but (i) the non-giant component infected in $L_B$ is connected with an interlayer edge (with expected value $e_B s_B$), (ii) this connecting edge transmits the epidemic (with probability $T_{BA}$), and (iii) the infection in $L_A$ results in an outbreak (with probability $S_A$). The probability of an outbreak in this scenario is therefore:

$$P_A = S_B + (1 - S_B)S_A e_B s_B T_{BA}. \tag{3}$$

### 3.2 Outbreak size in interconnected networks

While in monolayer networks the size and probability of an outbreak are the same ($S = P$), this is not the case in interconnected networks, as we show below. In scenario 1, the size of an outbreak in $L_A$, which we will denote as $R_\infty^A$, is augmented by the possible infection of the non-giant components in the percolation process on that layer. Since we assume an outbreak occurs in $L_A$, an outbreak must occur in $L_B$ as well. The giant-component in $L_B$ may be connected via interlayer edges to non-giant component in $L_A$ which may now become infected. The number of interlayer edges connected to the giant component in $L_B$ is $S_B e_B n_B$, of which $1 - S_A$ are connected to non-giant components in $L_A$. Each such infection in $L_A$ increases the size of the outbreak by $\frac{s_A}{n_A}$ (due to assumption 2). Therefore,

$$\begin{aligned} R_\infty^A &= S_A + S_B(1 - S_A)s_A e_B \frac{n_B}{n_A} T_{BA} \\ &= S_A + S_B(1 - S_A)s_A e_A T_{BA} \end{aligned} \quad (4)$$

Notice that equation 4 is different from both equations 2 and 3, and only transmission in the direction $L_B \to L_A$ plays a role.

For scenario 2 the formulation is similar. Since we assume that an outbreak occurs in $L_A$, it is of no significance to the outbreak size where the epidemic originated, since an outbreak must occur in $L_B$. Therefore, $L_B$ augments the outbreak in $L_A$ in a similar manner as for Scenario 1, and equation 4 holds in this case as well. Therefore (and in contrast to the probability of an epidemic), only transmission rates in the direction $L_A \to L_B$ affect outbreak size, regardless of the source of the epidemic.

## 4 Numeric simulations of the spreading process

To test the analytic solutions, we compared them with numeric simulations. We used the Erdős-Rényi (ER) model as it is commonly considered as a 'null' network structure and because for the purpose of theoretical work, which aims to provide a general framework to generate predictions, the use of well-defined network models is advantageous [34]. We find explicit solutions for the ER model (see the Supplementary Information for derivation of $S$ and $s$ for ER networks), and compare them to simulation results. Nonetheless, real animal or human networks do not necessarily fit into predefined models and our analytic framework can be used with any network structure.

Following [27, 28, 35], we generated two ER networks of size $n = 1000$ and a mean degree of $K = 10$, and connected them with $E = 1000$ interlayer edges. That is, on average, every node in layer A connects to one node in layer B ($e_A = e_B = 1$). Although in some systems it may be logical or necessary to connect nodes between layers non-uniformly (e.g., preferentially connecting between nodes with highest intralayer node degree), Wang & Xiao [36] have shown that such non-random interlayer connectivity leads to similar dynamics as with random connectivity. We therefore connected nodes randomly between $L_A$ and $L_B$.

In multihost systems, infection probability is usually assumed to be lower (or equal, at most) between species than within-species [13, 21]. We therefore considered $\beta_{AB}, \beta_{BA} \leq \beta_{AA} = \beta_{BB} = 0.03$. We tested for the effect of asymmetry in infection probability by varying $\beta_{AB}$ and $\beta_{BA}$ from 0 to 0.03 (including) in increments of 0.0005, resulting in $61 \times 61 = 3721$ $(\beta_{AB}, \beta_{BA})$ combinations. For each $(\beta_{AB}, \beta_{BA})$ combination, we randomly generated 100 interconnected networks, each of which was infected 1000 times, resulting in 100,000 simulations per combination. We set $\tau_A = \tau_B = 6$. Each simulation was run until there were no more infected individuals. In each simulation, we calculated $r_\infty$ — the final proportion of individuals in layer $L_A$ (the focal host) who were infected at some point during the simulation. We then calculated two properties for each $(\beta_{AB}, \beta_{BA})$ combination: (i) The probability of an outbreak, $\mathcal{P}_e$, defined as the proportion of simulations (out of 100,000) in which the pathogen infected more than 10% of the population. While the selection of 10% is arbitrary, the exact value does not change the results. This is due to the strong bi-modality in the infection process, which is a consequence of the two types of components—giant and non-giant—rather than a continous distribution of componenet sizes [31]. We compare $\mathcal{P}_e$ to $P_A$ calculated using the analytic solution. (ii) Outbreak size—the mean number of individuals infected in those simulations that have passed the 10% threshold, $\mathcal{R}_\infty$. We compare $\mathcal{R}_\infty$ to $R_\infty^A$ calculated using the analytic solution. Note that we use normal letters for notation of analytic parameters (e.g., $P_A$) and calligraphic letters for notation of parameters related to the simulations (e.g., $\mathcal{P}_e$).

Our choice of parameter values for the numerical simulations was made to facilitate the way we illustrate the analytic approach (we discuss the divergence of numerical simulations from analytic solutions below). The analytic framework is flexible enough to allow for any choice of parameters which satisfy our assumptions. For example, one could choose unequal infection probabilities or recovery rates for the two layers or other range of values for $\beta_{AB}, \beta_{BA}$.

### 4.1 Effect on the probability of an outbreak

Equations 2 and 3 predict that the probability of an outbreak in the focal species, $P_A$, will strongly depend on the source of infection. This is so because these two equations differ in their parameters. Accordingly, we find in the simulations that when the disease originates in the focal host $(L_A)$, $\beta_{AB}$ determines $\mathcal{P}_e$ to a large extent (Fig. 2a), whereas when the disease originates in $L_B$, $\mathcal{P}_e$ is determined almost exclusively by $\beta_{BA}$ (Fig. 2b). The reversal of the roles of $\beta_{BA}$ and $\beta_{AB}$ with different source of infections is illustrated by observing horizontal and vertical 'cross-sections' across the $\beta_{AB} - \beta_{BA}$ space (Fig. 3). It is evident from Fig. 3a that $\mathcal{P}_e$ increases approximately linearly with increasing $\beta_{AB}$, whereas it remains rather constant with changes in $\beta_{BA}$. The opposite pattern is evident in Fig. 3b, where $\mathcal{P}_e$ increases approximately linearly with an increase in $\beta_{BA}$ but almost does not change with $\beta_{AB}$.

The emergence of this behaviour is explained as follows. If an outbreak did not occur in the source layer (which can be either $L_A$ or $L_B$), then there must be transmission from the source to the non-source layer (the terms on the right in equations 2 and 3) for an outbreak to still occur in $L_A$. This transmission is a function of the probability of infection from the source layer to the non-source layer. Once the infection traversed between layers, there are two options: (i) there is no outbreak in the non-source layer and the epidemic dies out (re-transmission back to the source layer is negligible due to assumption 2). (ii) There is an outbreak and transmission back to the source layer is almost certain. In this case the infection probability from the non-source to the source layer plays no role. Hence, both the source of infection and the asymmetry in interlayer transmission determine the behaviour of $\mathcal{P}_e$.

### 4.2  Effect on outbreak size

In contrast to the marked qualitative effect of the source of infection on $\mathcal{P}_e$, it does not seem to have any major qualitative effect on the behaviour of $\mathcal{R}_\infty$ (Fig. 2c,d and Fig. 3c,d). The cross-sections show that $\mathcal{R}_\infty$ increases approximately linearly as $\beta_{BA}$ increases but remains almost unaffected by $\beta_{AB}$, regardless of the source of infection. This is so because $\mathcal{R}_\infty$ is conditioned on an outbreak already occurring in $L_A$, and hence also in $L_B$. The additional increase in $\mathcal{R}_\infty$ (compared to the case where it is a monolayer network) is due to infections from the giant component in $L_B$ to non-giant components in $L_A$ (again, under our assumptions back-transmission from non-giant components in negligible). Therefore, the source of infection has no affect on $\mathcal{R}_\infty$. In addition, $\beta_{AB}$ has no effect on $\mathcal{R}_\infty$ as evident from the line which is parallel to the x-axis in Fig. 3 (in red). This is in accordance with equation 4 which, as noted above, does not include transmission in the direction $L_A \to L_B$.

### 4.3  Limits of the analytic formulation

Overall, it is evident from Fig. 3 that the numeric simulations provide general support for the analytic solutions; however, there are several discrepancies which deserve attention and which also give us some insight into the infection process. First, for very low values of interlayer infection probabilities, the analytic solutions significantly diverge from the numerical simulations. In scenario 2 (Fig. 3b,d), we observe that for the $\beta_{BA}$ cross-section (blue colour in the figure) infection probability determines epidemic dynamics in two ways: (i) a minimal $\beta_{BA}$ is needed for any outbreak to occur in $L_A$, and (ii) once this minimal threshold is exceeded, $\beta_{BA}$ significantly contributes to the probability and size of outbreaks. We observe a sharp decline at low $\beta_{BA}$ values, which does not conform with the analytic solutions. This is because in the analytic formulation we assume sufficient transmission (assumption 1), which is a condition not met for low parameter values. Once no $L_B \to L_A$ transmission occurs, no infection occurs in the focal layer.

In scenario 1, sharp declines in outbreak probability and size are observed for low $\beta_{BA}$ (Fig. 3a) and $\beta_{AB}$, respectively (Fig. 3c). In this scenario, outbreak size and probability in $L_A$ are above zero independently of interlayer transmission, since the source of the epidemic is in $L_A$ itself. Therefore, at very low $\beta_{BA}$ and $\beta_{AB}$ values outbreak probability and size have non-zero baseline values. These baseline values are the non-zero intercepts with the y-axis, and correspond to the expected dynamics in $L_A$ as if it were a monolayer network (because there is either no $L_A \to L_B$ or no $L_B \to L_A$ transmission). For any additional contribution from $L_B$, both $\beta_{BA}$ and $\beta_{AB}$ must be non-zero, as disease must transmit back and forth between the layers to affect $L_A$. Due to assumption 1 (a minimal amount of disease flow between layers), we see that the analytic formulations may overestimate disease risk at low transmission rates.

A second type of discrepancy is seen at high interlayer transmission rates for either $\beta_{BA}$ in Fig. 3a (red) or for $\beta_{AB}$ in Fig. 3b,c,d (blue). Here, the analytic formulations predict a linear relation while the numeric simulations seem to point to a sub-linear one. One possible explanation for this observation concerns the expected contribution of the interlayer edges for transmission. The analytic formulations assume that the expected contribution of each transmitting interlayer edge is additive, since only one transmitting edge connects non-giant to giant components (assumption 2), leading to a linear contribution of the interlayer probabilities to outbreak probability and size. However, when transmissibility is high enough, some interlayer edges which transmit the disease may connect to the same non-giant component, violating the assumption of additive contribution to outbreak size and probability, resulting in sublinear contribution to infection in $L_A$.

## 5 Discussion

In this paper we have applied analytic and numeric analysis to investigate the outcomes of disease spread in interconnected networks under asymmetric infection probabilities. Our approach was motivated by the ecology and epidemiology of multihost diseases and we tracked disease outcomes in a focal host while considering the influence of a second host. We have shown that asymmetry in interlayer infection probabilities and the source of infection are both important factors in determining disease dynamics. For example, outbreak probability and size were affected differently by the source of infection, emphasizing the importance of considering different measures of disease risk, and a more detailed understanding of the mechanisms that underlie them. We discuss our results in the context of disease ecology and multilayer network theory.

We find that the dynamics in the focal host depends on whether the disease originated in this host itself or in the non-focal host. Many models of multihost pathogen transmission in disease ecology consider a *target* species infected by a reservoir *source* (disease originates in $L_B$ and $\beta_{AB} \to 0$; e.g., [13, 22]). In the simulations, we assumed a constant maximum within-species infection probability and thus examine the continuum between emerging infectious diseases such as influenza ($\beta_{BA} \approx 0$) and true multihost diseases such as foot-and-mouth disease ($\beta_{AB} \approx \beta_{BA} \approx \beta_{AA} \approx \beta_{BB}$) [13, 25]. On this continuum (and when the disease originates in the non-focal host), our results are consistent with previous studies as we show that the role of $\beta_{AB}$ is minor in determining outbreak probability and size (see horizontal cross-sections in Fig. 3).

Note that in the case where between-host and within-host infection probabilities are equal (a true multihost pathogen; [13]), dynamics will be determined only by the distribution of interlayer and intralayer edges. Dynamics in this class of network topology has been explored thoroughly in the context of network community structure (see [3] for a review), but represent a particular case of the more general epidemic behaviour we investigate here, which is more relevant for disease ecology.

While one-way transmission models are particularly relevant for zoonotic diseases, where identifying the source of infection has been a major endeavour [18], the distinction between source and target hosts is blurred for many parasites which can switch between host species. For example, cowpox virus can infect both bank voles (*Clethrionomys glareolus*) and wood mice (*Apodemus sylvaticus*), with different infection probabilities [37] and humans and apes share several pathogens [38]. Indeed, any infection probability is unique to a host-pathogen combination and a result of ecological, epidemiological and evolutionary processes [39–42]. Therefore, there are several host traits which cause among-host heterogeneity in infection probability to a given parasite [43]. Accordingly, our results show that considering between-host infection probabilities in both directions is crucial for disease outcomes. In addition, the qualitative difference in behaviour of outbreak size and probability (in relation to the source of infection) highlights the need to include two-way infections and consider the source of infection in future models. It also has some empirical support. For example, work on the shift of *Microbotryum violaceum*, the causal agent of anther-smut disease, between plant species of the genus *Silene* has shown that the transmission of the pathogen in both directions between the new and old plant hosts increases disease prevalence in both host plants [44], supporting our theoretical results.

Regardless of the host of origin, the non-focal host serves as an amplifier of the disease, in the sense that an increase in species diversity increases disease risk (here measured as outbreak probability and size) [23]. This is clearly evident from equations 2, 3 and 4 which have two additive terms. However, the contribution of the non-focal host to epidemic size depends mainly on the transmissibility in the direction $L_B \to L_A$. Another way in which the non-focal host functions as a disease amplifier is that epidemiological cases in the focal host are not necessarily linked. This is because in networks, unlike in mean-field models, there are alternative routes for the infection to spread by carrying the disease to parts of the focal host network where it has not reached or would not have reached by infection within the focal host alone. This phenomenon becomes more likely with increasing between-species transmission.

Our study also has implications for the more general field of network science. Previous theoretical studies of epidemic spread in interconnected networks have focused on topological patterns that enable disease establishment (i.e., $R_0 > 1$), while assuming symmetry in interlayer infection probabilities [5–7, 27, 28, 36]. Only one study that we know of has addressed asymmetric rates of infection between two networks [35] but the authors investigated an analytic solution to the combination of infection probabilities that allows crossing the epidemic threshold ($R_0 > 1$), and not the explicit effect of asymmetry on the disease outcomes. Our study advances the theory of diffusion in interconnected networks as it emphasizes that asymmetry in interlayer infection probabilities can change the expected diffusion dynamics due to a joint effect with the source of infection on epidemic size and probability.

## Conclusions

We find that epidemic size and probability, two proxies for disease risk, depend on particular configurations of two-way transmission between host networks. However, models of disease dynamics in disease ecology usually consider only one-way transmission. Hence, we believe that the study of interconnected multihost network systems will contribute to a better understanding of disease dynamics in multihost systems on the one hand and advance the theoretical understanding of epidemic spread in multilayer network on the other. Unfortunately, it is difficult to obtain good data which include both within- and between-population contacts as well as information on infection probabilities. Such data is needed to corroborate theoretical results, and we emphasize the importance of data collection and the generation of multi-species data sets.

We suggest that future models investigate two-way disease transmission which involves asymmetric infection probabilities. This is very relevant for diseases that affect non-human organisms but we should also remember that pathogens can also be transmitted from humans to other hosts and that in some diseases several hosts are involved in addition to humans [17]. Moreover, identifying the source of infection is crucial. For example, if the pathogen is endemic in a focal host but infections are usually, (but not necessarily) sub-clinical, then a disease can originate in the focal host itself and then be amplified by a reservoir.

Finally, our results also have consequences for spreading processes in non-disease systems. For example, in social multilayer networks information can flow among different layers (e.g., between different online social networks) in a non-symmetric fashion. Our study is therefore relevant for the general theory of diffusion in multilayer networks and it will be valuable to explore additional consequences of asymmetry in interlayer diffusion probabilities.

## Acknowledgements

We thank Monika Böhm for helpful comments on previous drafts of this manuscript. This is publication no. XXX of the Mitrani Department of Desert Ecology.

## Code Accessibility

We provide the complete code (DOI:10.6084/m9.figshare.2058879) in the Figshare online repository.

## Competing Interests

We have no competing interests.

## Authors' Contributions

SP conceived the idea and wrote the manuscript. SP and GG designed the study. GG developed analytic formulations and commented on the manuscript. BK contributed ideas and commented on the manuscript. YZ implemented simulations and commented on the manuscript. SP, GG and YZ analysed the results. All authors gave final approval for publication.

## Funding

SP was supported by a James S. McDonnell foundation postdoctoral fellowship for the study of complex systems and by a Fulbright fellowship sponsored by the U.S. Department of State.

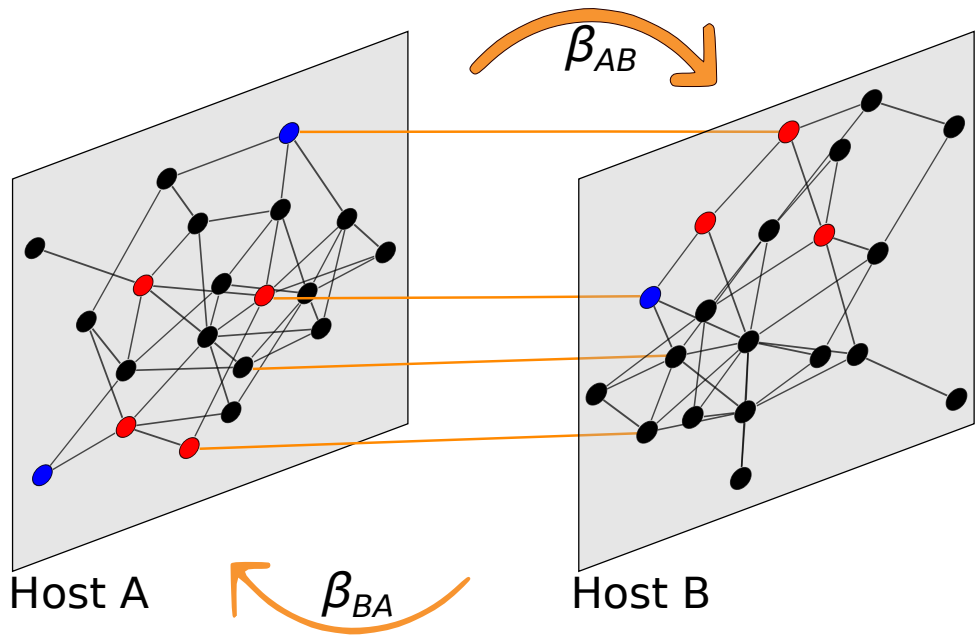

**Fig. 1 A multilayer interconnected network between hosts A (left side) and B (right side).** Intralayer edges (in black) represent contacts between individuals of the same host species. Interlayer edges (in orange) represent contacts between individuals of different host species. Infection probabilities between hosts A and B ($\beta_{AB}$ and $\beta_{BA}$) may be asymmetric (represented by different width of orange arrows). Individuals can be susceptible (black nodes), infected (red nodes) or recovered (blue nodes).

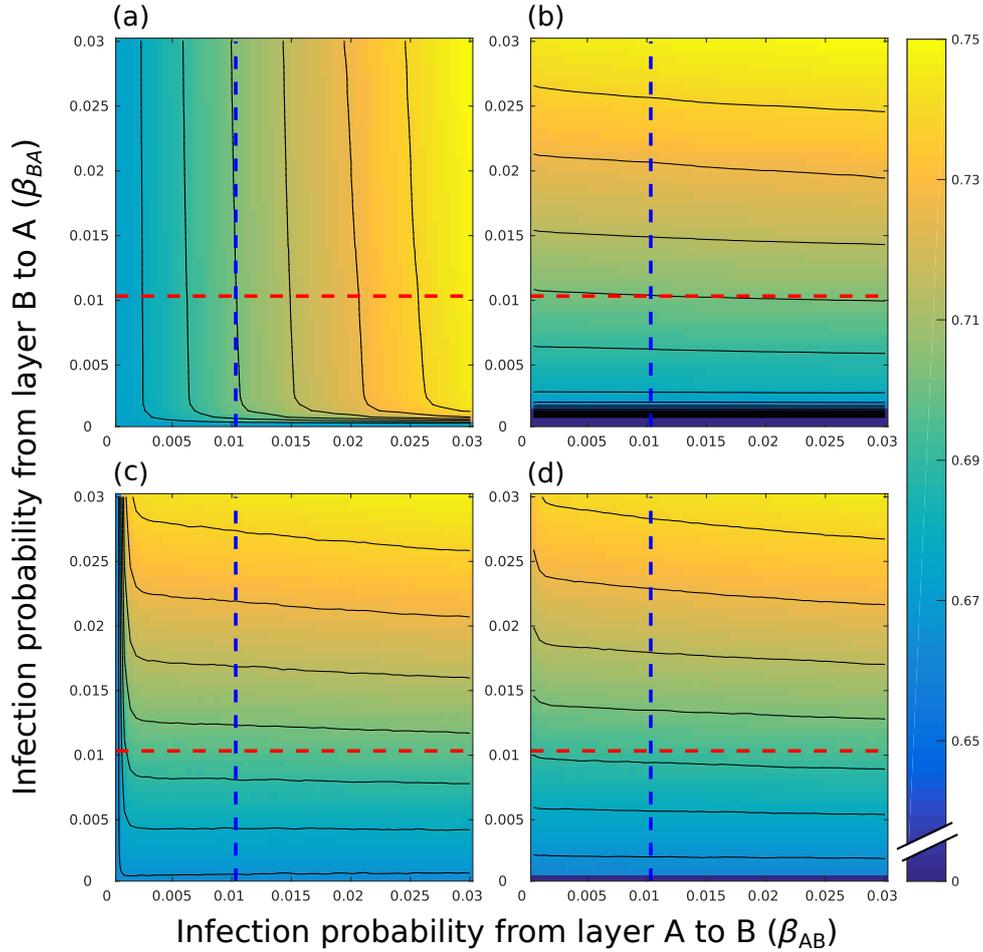

**Fig. 2 Disease dynamics as a function of source of infection and inter-layer infection probabilities.** Panels (a) and (b) depict the probability that a disease would infect at least 10% of the individuals in $L_A$ ($\mathcal{P}_e$), when the disease originates in $L_A$ and $L_B$, respectively. Panels (c) and (d) depict the mean number of individuals infected in those simulations that have passed the 10% threshold ($\mathcal{R}_\infty$) when the disease originates in $L_A$ and $L_B$, respectively. The values of $\mathcal{P}_e$ and $\mathcal{R}_\infty$ in the plots are a function of the infection probabilities $\beta_{AB}$ and $\beta_{BA}$. Note that selecting a threshold other than 10% does not change the results.

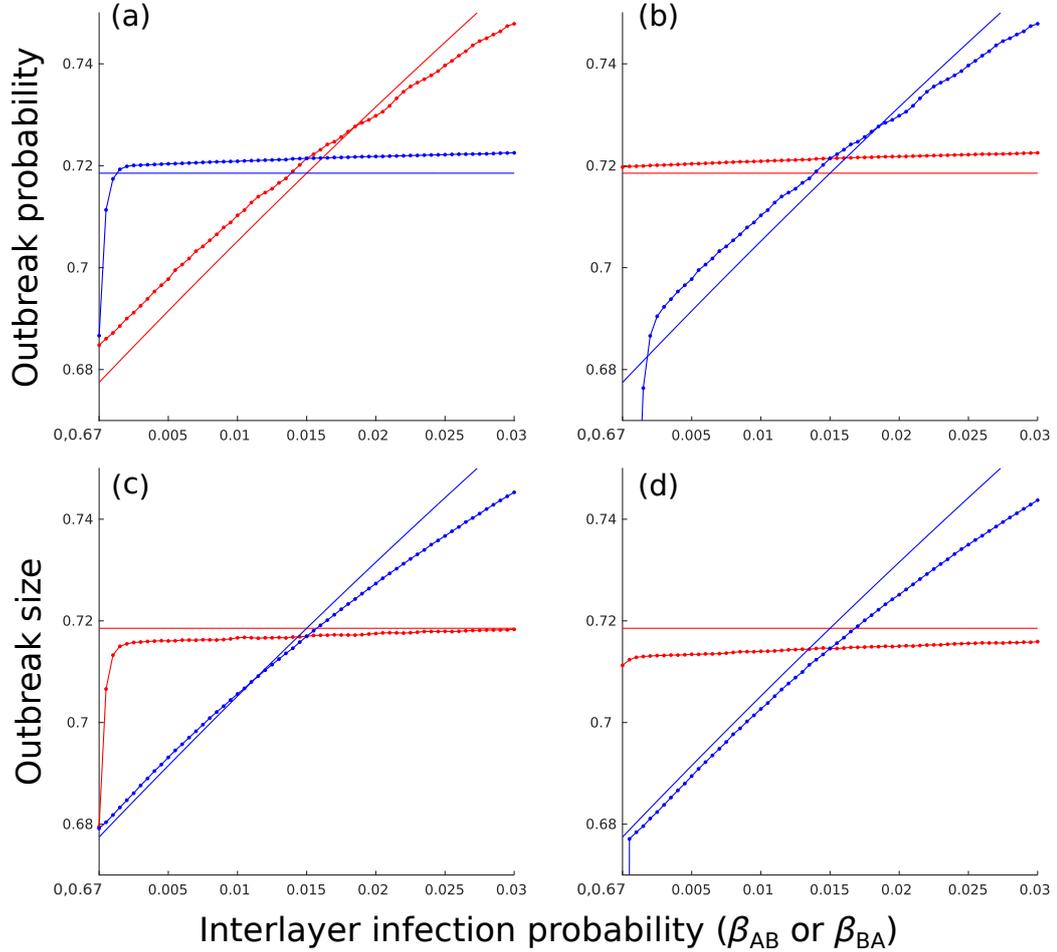

**Fig. 3 Comparison of analytic and simulation results**. Red and blue colours represent cross-sections across $\beta_{BA} = 0.01$ (red) and $\beta_{AB} = 0.01$ (blue) in the $\beta_{AB} - \beta_{BA}$ parameter space from Figure 2. Solid lines are analytic solutions and closed circles are simulation results. Panels (a) and (b) depict the probability of an outbreak in $L_A$ (denoted as $P_A$ for analytic solutions and $\mathcal{P}_e$ for simulations in the main text). Panels (c) and (d) depict the mean outbreak size in $L_A$ (denoted as $R_\infty^A$ for analytic solutions and $\mathcal{R}_\infty$ for simulations in the main text). In panels (a) and (c) the disease originates in $L_A$ (scenario 1), whereas in panels (b) and (d) it originates in $L_B$ (scenario 2). Analytic solutions are shown in equations 2, 3, 4 and the Supplementary Information. In (b) and (d) the Y-intercept of the numeric results for $\beta_{AB} = 0.01$ (blue) is at zero (not shown for clarity).

## Supplementary Information

### Size of giant and non-giant components under percolation process in arbitrary random networks in the supercritical phase

In this supplement we present the formulation of the expected size of giant components and expected mean size of non-giant components in random networks following bond percolation in the network, in the supercritical regime. These sizes correspond to the expected extent of SIR epidemics (above the percolation threshold) in the network, either when an outbreak occurs (giant component) or when it doesn't (non-giant component). The derivation of the size of the giant component was shown by Newman [31] and we reproduce the result here in the context of our work. We also show the derivation of the mean size of the non-giant component using methods similar to those employed in [31].

We consider a random network with degree distribution $D$, for which the mean degree is $z = <D>$; we will denote the generating function of $D$ as $G_0(x)$. We now select an edge at random, and select randomly one node incident to that edge, and count the number of neighbours of that node, excluding the node connected to the edge we selected initialy. As shown by Newman [31], the distribution of these counts is

$$Q(x = k) = \frac{(k+1)D(x = k+1)}{\sum_j j D(x = j)} \quad \text{(SI.1)}$$

with a mean of $<Q> = \frac{<D^2> - <D>}{<D>}$; thus, the average number of nodes two steps away from a randomly selected node is $z_2 = <D^2> - <D>$. We denote the generating function of $Q$ as $G_1(x)$. By definition

$$G_1(x) = \sum_{k=0}^{\infty} Q(x = k) x^k = \frac{G_0'(x)}{z} \quad \text{(SI.2)}$$

**Sizes of giant and non-giant components in random networks**

First we find the distribution of non-giant components in networks without a giant component. These are very small components with only a few nodes, and it can be safely assumed (unless the network is extremely dense) that these components do not include loops (i.e., they are tree-like). We now choose a random edge and a node incident to it as before, and count the neighbors of this node excluding the node connected to the initially selected edge (so it is the size of the neighborhood minus one). Let $H_1(x)$ be the generating function of the distribution of sizes of these components. Since we can continue to follow edges in this ways and sum the number of reachable nodes at each level, as there are no self-loops, we can define the generating function recursively:

$$H_1(x) = \sum_{k=0}^{\infty} Q(x=k)[H_1(x)]^k = xG_1(H_1(x)) \qquad (SI.3)$$

Here we use the fact that the sum of $n$ independent random variables with the same distribution can be found by observing the $n^{th}$ power of the generating function of the distribution. The distribution of component sizes in this case, $H_0(x)$, is found by selecting a node at random and quantifying the size of the component of that node by following the edges connected to the node, which have distribution $D$:

$$H_0(x) = \sum_{k=0}^{\infty} D(x=k)[H_1(x)]^k = xG_0(H_1(x)) \qquad (SI.4)$$

Solving equations SI.3 and SI.4 gives us the generating function of component sizes when no giant component exists, and the mean component size can then be found by evaluating the derivative of the generating function $H_0'(x)$ at $x = 1$.

When a giant component does exist (which is the scenario of interest in our case), evaluating $H_0(x)$ at $x = 1$ will give us the total fraction of the network occupied by non-giant components. Therefore, the size of the giant component is simply

$$S = 1 - H_0(1) \tag{SI.5}$$

To find the mean size of the non-giant components we must normalize the derivative of the generating function with the fraction of the network occupied by the non-giant components, $H_0(1)$:

$$s = \frac{H_0'(1)}{H_0(1)} \tag{SI.6}$$

**Size of giant component under bond percolation**

Under bond percolation with transmissiblity $T$, we simply remove edges from the network with probability $1 - T$ and evaluate the sizes of the component in the resulting network. Thus, the degree distributions need to be reformulated to generate the correct generating functions equivalent to equations SI.3 and SI.4. Since the number of edges remaining connected for each node is binomaly distributed with parameter $T$, the new generating functions are [45]:

$$G_0^T(x) = \sum_{m=0}^{\infty}\sum_{k=m}^{\infty} D(x=k)\binom{k}{m}T^m(1-T)^{k-m}x^m = G_0(1+(x-1)T) \tag{SI.7}$$

and similarly

$$G_1^T(x) = G_1(1+(x-1)T). \tag{SI.8}$$

Exchanging $G_0(x)$ and $G_1(x)$ for $G_0^T(x)$ and $G_1^T(x)$ in equations SI.3 and SI.4 results in the following generating functions:

$$H_1^T(x) = 1 - T + TxG_1(H_1^T(x)) \tag{SI.9}$$

and

$$H_0^T(x) = xG_0(H_1^T(x)) \tag{SI.10}$$

The size of the giant component, $S$, under the bond percolation process, which is the size of an outbreak if an outbreak occurs, can be found by solving equations SI.5 and SI.9 ($v = H_1^T(1)$):

$$S = 1 - G_0(v) \quad \text{and} \quad v = 1 - T + TG_1(v) \tag{SI.11}$$

**Mean size of non-giant components under bond percolation**

We will proceed to the derivation of the mean size of the non-giant components along similar lines. From equation SI.6 we have that the mean size of the giant components is

$$s = \frac{H_0^{T'}(1)}{H_0^T(1)} \tag{SI.12}$$

For the nominator we have (equation SI.10)

$$H_0^T(1) = G_0(H_1^T(1)) = G_0(v) \tag{SI.13}$$

with the same $v$ found in the solution to equation SI.11. For the denominator we find the derivative:

$$H_0^{T'}(x) = G_0(H_1^T(x)) + xH_1^{T'}(x)G_0'(H_1^T(x)) \tag{SI.14}$$

and so

$$H_0^{T'}(1) = G_0(v) + H_1^{T'}(1)G_0'(v) \tag{SI.15}$$

To find $H_1^{T'}(x)$ we take the derivative of equation SI.9:

$$H_1^{T'}(x) = TG_1\big(H_1^T(x)\big) + TxH_1^{T'}(x)G_1'\big(H_1^T(x)\big) \tag{SI.16}$$

Now for $x = 1$

$$\begin{aligned}H_1^{T'}(1) &= TG_1\big(H_1^T(1)\big) + TH_1^{T'}(1)G_1'\big(H_1^T(1)\big) \\ &= TG_1(v) + TH_1^{T'}(1)G_1'(v)\end{aligned} \tag{SI.17}$$

Finally we can find $H_1^{T'}(1)$ by solving $(u = H_1^{T'}(1))$:

$$u = TG_1(v) + TuG_1'(v) \tag{SI.18}$$

From equations SI.12, SI.13 and SI.15 (with the solution to SI.18) we obtain the mean size of the non-giant components

$$s = \frac{G_0(v) + uG_0'(v)}{G_0(v)} = 1 + \frac{uG_0'(v)}{G_0(v)} = 1 + \frac{uzG_1(v)}{1-S} \tag{SI.19}$$

**Explicit solutions for component sizes in Erdős-Rényi networks**

Solutions for equations SI.11 and SI.19 are generally found using numerical methods, but for the Erdős-Rényi network we can find explicit solutions. These explicit solutions are used for comparison with the simulation analysis in the main text, and are shown in Fig. 3. Derivations of these solutions from the formulation above is straightforward. In the Erdős-Rényi random networks the degree distribution follows a Poisson distribution with parameter $z$ (the mean degree). Thus, the generating function of the degree distribution in this case is $G_0(x) = e^{z(x-1)}$. Substituting in equation SI.2 we find that $G_1(x) = e^{z(x-1)}$. From equation SI.11

$$v = 1 - T + Te^{z(v-1)} \quad \text{and} \quad S = 1 - e^{z(v-1)} \tag{SI.20}$$

The solution to this equation is

$$S = 1 - e^{z(1-T-\frac{W_0(-Tze^{-Tz})}{z}-1)} \tag{SI.21}$$

where $W_0$ is the main branch of the Lambert W function. We can now solve $u$ using equation SI.18:

$$u = -\frac{1}{z\left(1 + \frac{1}{W_0(-Tze^{-Tz})}\right)} \tag{SI.22}$$

The mean size of the giant component can be found by substituting $u$ in equation SI.19 and using $v$ from the solution to equation SI.20:

$$s = 1 - \frac{G_1(v)}{(1-S)z\left(1 + \frac{1}{W_0(-Tze^{-Tz})}\right)} \tag{SI.23}$$

Therefore, the probability of an outbreak occurring in an Erdős-Rényi network with mean degree $z$ from a single infected individual is the probability for belonging to the giant component, $S$ (equation SI.21), and the expected size of the outbreak is also $S$. If an outbreak does not occur, it is expected that only $s$ (equation SI.23) individuals will become infected.